\documentclass{emulateapj}
\lefthead{GOULD, GAUDI \& HAN} \righthead{DETECTION OF EARTH-MASS PLANETS}
\def\kpc{{\rm kpc}}

\def\au{{\rm AU}}
\def\min{{\rm min}}
\def\bol{{\rm bol}}
\def\muas{{\mu \rm as}}
\def\sn{{\rm S/N}}
\def\me{M_\oplus}

\begin{document}

\title{Prospects for the Detection of Earth-Mass Planets}
\author 
{Andrew Gould\altaffilmark{1}, B.\ Scott Gaudi\altaffilmark{2}, and
Cheongho Han\altaffilmark{1,3}}
\altaffiltext{1}{Ohio State University, Department of Astronomy, Columbus, OH 43210}
\altaffiltext{2}{Harvard-Smithsonian Center for Astrophysics, 60 Garden St., Cambridge, MA 02138}
\altaffiltext{3}{Department of Physics, Institute for Basic Science
Research, Chungbuk National University, Chongju 361-763, Korea}
\email{gould@astronomy.ohio-state.edu, sgaudi@cfa.harvard.edu,cheongho@astroph.chungbuk.ac.kr}

\doublespace

\begin{abstract}
We compare potential state-of-the-art experiments for detecting
Earth-mass planets around main-sequence stars using radial velocities,
transits, astrometry, and microlensing.  For conventionally-discussed
signal-to-noise ratio ($\sn$) thresholds, $\sn \sim 8$, the last three
methods are roughly comparable in terms of both the total number of
planets detected and the mass distribution of their host stars.
However we argue that $\sn \sim 25$ is
a more conservative and realistic $\sn$ threshold. 
We show analytically
and numerically that the decline in the number of detections as a
function of $\sn$ is very steep for radial velocities, transits, and
astrometry, such that the number of expected detections at $\sn \sim
25$ is more than an order-of-magnitude smaller than at conventional
$\sn$ thresholds.  Indeed, unless Earth-mass planets are very common
or are packed much closer to their parent stars than in the solar
system, future searches using these methods (as they
are currently planned) may not yield any reliable
Earth-mass planet detections.  On the
other hand, microlensing has a much shallower $\sn$ slope than the
other techniques and so has much greater sensitivity at realistic
$\sn$ thresholds.  We show that even if all stars have Earth-mass
planets at periods of one year (and adopting other optimistic assumptions 
as well), the combined yield of all four
techniques would be the detection of only about five such planets at
$\sn \sim 25$.
\end{abstract}
\keywords{astrobiology -- astrometry -- 
planetary systems -- extraterrestrial intelligence}
\clearpage
 
\section{Introduction}\label{sec:intro}

The detection of terrestrial planets around main-sequence stars
represents a long-standing major goal of observational astronomy that
may finally be realized within the next decade.
Terrestrial (i.e.\ rocky) planets are interesting for several reasons.
The frequency, mass, and orbital distribution of planets with masses
of $\sim1-10$ Earth masses ($\me$) yield strong constraints on the physical
mechanisms at work in planetary formation and would provide a crucial
observational test of planet formation theories.  When combined with
similar distributions of the properties of more massive planets,
this information would
allow for stringent tests of core-accretion models for the formation
of gas and ice-giant planets (e.g.\ \citealt{il04}).  Terrestrial
planets are also commonly assumed to be the most favorable places to
look for extrasolar life.  Especially favored are planets with
separations from their parent star that place them in the so-called
``habitable zone,'' roughly the range of distances at which liquid water
can exist, although it is important to keep in mind that not all
low-mass planets in the habitable zone are necessarily terrestrial
\citep{kuchner03,raymond04}, nor should gas-giant planets in the habitable zone
be disregarded, as their satellites may well be capable of supporting
life \citep{wkw97}.

An accurate and robust determination of $\eta_\oplus$, the frequency
of Earth-mass planets around main-sequence stars, is also essential
for the design of future missions that aim to directly image and take
spectra of terrestrial planets around the nearest stars, such as the
{\it Terrestrial Planet
Finder}\footnote{http://planetquest.jpl.nasa.gov/TPF/tpf\_index.html}
and {\it Darwin}\footnote{http://www.esa.int/science/darwin}.  In many
design concepts for these missions, the required diameter $D$ of the
primary collecting mirror is directly proportional to the distance $d$
of the stars to be surveyed \citep{beichman04}.  Thus for a fixed
number of planets in the survey volume, $D \propto
\eta_{\oplus}^{-1/3}$.  Since the cost and feasibility of such
missions is likely to be an extremely strong function of $D$, accurate
determination of $\eta_{\oplus}$ is essential \citep{beichman03}.

Several different indirect detection techniques have the potential to
determine the frequency of terrestrial planets around
solar-type stars.  Transiting terrestrial planets are detectable
photometrically via the periodic dimming of their parent star.
For a planet with the radius of the Earth orbiting a main-sequence
star, the fractional change in brightness during the transit is
$O(10^{-4})$; achieving photometric precisions at this level or better
generally requires observations from space (but see \citealt{gpd}).
Also, the low transit probability and duty cycle requires that many
stars must be monitored continuously.  The space mission {\it
Kepler}\footnote{http://www.kepler.arc.nasa.gov/} will continuously
monitor $\sim 10^5$ stars for four years in order to search for
transiting planets and will be sensitive to terrestrial planets
orbiting in the habitable zones of disk main-sequence stars at
distances of $d\sim 1~{\rm kpc}$.  Terrestrial planets can also be
detected from the astrometric wobble they induce on their parent star.
For planets of mass $\sim M_\oplus$ orbiting at a few $\au$ from
parent stars located at distances of a few parsecs, the induced
displacement is $O(\muas)$.  The {\it Space Interferometry Mission
(SIM)}\footnote{http://planetquest.jpl.nasa.gov/SIM/sim\_index.html}
will use ultra-high precision astrometry to search for planets in the
solar neighborhood ($d \la 20~{\rm pc}$), and will be sensitive to planets of a few
Earth-masses with orbits of a few $\au$ around the most nearby stars
\citep{sozzetti02,ft03,gff03}.  Terrestrial planets can be detected via
microlensing from the pronounced but brief deviation they induce on
microlensing events of their parent stars \citep{mp91,gl92}.
Microlensing is potentially very sensitive to Earth-mass planets
\citep{eplan}. However, the rarity and brevity of the signal induced by
the planets requires the continuous monitoring of a large number of
stars.  The proposed space mission {\it Microlensing Planet Finder (MPF)} \citep{gest} would search
for planets via microlensing by continuously monitoring $\sim 10^8$
main-sequence stars toward the Galactic bulge.  This mission would be
sensitive to terrestrial planets with few AU orbits around Galactic disk 
and bulge main-sequence stars at distances of $1-10~\kpc$.
Terrestrial planets may also be detectable from the ground with
microlensing \citep{bennett04,ghg}. Orbiting planets induce radial velocity (RV) variations
on their parent stars: an Earth-mass planet orbiting a solar-type star
with a period of $0.5~{\rm yr}$ induces a RV variation on its parent star of
amplitude $\sim 0.1~{\rm m~s^{-1}}$.  This is well below the
systematic limit of $1~{\rm m~s^{-1}}$ typically quoted for current RV
surveys. However, it is not clear whether this represents a hard floor, or
whether the systematic errors can be averaged out over repeated
observations.  Therefore, it may also be possible to detect Earth-mass
planets from the ground with RV.

Although extrasolar Earth-mass planets have yet to be discovered orbiting main-sequence
stars, they have in fact already been discovered around a neutron star using
pulsar timing \citep{wf92,wolz,kw03}.  However, subsequent pulsar searches have
shown pulsar planets to be a very rare phenomenon.  Furthermore, 
such planets probably formed after the creation of the 
pulsar \citep{hansen02} and are unlikely to be suitable for life.

All of the proposed detection methods discussed above
are difficult and require ambitious,
cutting-edge experiments.  They are therefore both quite expensive and
complicated, and thus they require very careful and conservative 
justification that is based not only on their proposed scientific return, but
also on their feasibility and probability of success.  To date,
discussion has therefore focused on two key issues.
First, what information will be learned about such planets if they are
detected?  Second, how many planets can be detected down to a given
threshold in signal-to-noise ratio $\sn$?  However, there is a
third dimension to this discussion that has mostly been ignored: 
what is the expected distribution of $\sn$ with which planets are detected?

At first sight, this appears to be a rather arcane question.  In fact, 
it is crucial.  We will argue that, in sharp
contrast to experiments that can easily be repeated or that return null
results, difficult-to-repeat astronomical experiments that purport to 
detect novel phenomena must have a high $\sn$ threshold to be believable.
Or, what amounts effectively to the same thing, a substantial fraction
of the detections must be at high $\sn$, thereby allowing one to 
characterize the believability of the detections at lower $\sn$.
This rule of thumb is unconsciously followed by all observers, or at
least all observers who develop respectable track records over the
long term.  It should also be applied to the daunting task of
characterizing Earth-mass planets.

Here we quantify the sensitivity of future experiments
to Earth-mass planets using the four proposed techniques: 
transits, astrometry, RV, and microlensing.  We begin by 
discussing the complementarity
of these four techniques, both in terms of the physical parameters measured and the orbital separations of the planets (\S\ref{sec:complement}).  We then
state the methods and assumptions used in evaluating the sensitivity of these techniques (\S\ref{sec:methods}).  We next discuss
the expected form of the $\sn$ distribution for each of the four techniques (\S\ref{sec:scaling}).  We present a number of arguments suggesting that conventional $\sn$ thresholds are inadequate, and as a result that the
number of reliable detections is likely to be considerably
smaller than currently envisioned (\S\ref{sec:thresholds}).  We then explore the
number of expected planet detections from  each technique (\S\ref{sec:results})  as a function of the signal-to-noise threshold (\S\ref{sec:numvthresh}), the primary mass (\S\ref{sec:comp}), 
planet period (\S\ref{sec:period}), and planet mass (\S\ref{sec:mass}).  We conclude with a discussion of the implications of our results (\S\ref{sec:discuss}).

\section{Complementarity of Proposed Techniques}\label{sec:complement}

Different search techniques are intrinsically sensitive to terrestrial
planets with different characteristics.  In terms of orbital
separation, astrometric
sensitivity scales as the planet semi-major axis, $a$, while for RV,
the sensitivity is $\propto a^{-1/2}$.  The sensitivity of a
magnitude-limited transit survey is $\propto a^{-5/2}$, but this
technique is particularly prized for its ability to detect terrestrial
planets in the habitable zone.  Microlensing is most sensitive to
planets near the Einstein ring of the primary star, which for typical
lensing geometries lies at about $a\sim 2.5\,\au$.  These methods
are also complementary in the sense that they measure different physical
properties of the planets.  Transits provide
planet radii, while astrometric measurements provide planet masses.  Radial velocity
measurements yield the mass of the planet modulo the sine of the inclination of the planetary orbit.  In its primitive incarnation, microlensing routinely measures only the mass ratio between the planet
and star, although for ambitious next-generation surveys, the
mass of the planet will be measurable for a large fraction of the detections \citep{gest,mulensmass,hanetal04}.   Each method also has other unique advantages.  For example, only astrometry and RV routinely probe the architecture of
multiple-planet systems, whereas only
microlensing can detect Earth-mass planets at very large radii, or
free-floating Earth-mass planets that have been ejected from their systems
\citep{gest,hanetal04}.

These differences in sensitivity are basically a good thing: they
imply that a coordinated attack on this difficult problem from several
directions can substantially increase the range of planet parameters
and properties probed.  However, while beneficial to science, these
differences make it difficult to directly compare the overall sensitivity of
different techniques to the detection of terrestrial planets.  
Fortunately, the main result of this paper, a comparison
of the {\it relative} distribution functions of $\sn$, is generally
unaffected by exactly how the comparison is carried out.  Rather, this
affects only the {\it normalizations} of these distribution functions.

\section{Methods and Assumptions}\label{sec:methods}

In this section we provide our methods and assumptions used in
calculating the number of terrestrial planets detectable by each of
the four techniques considered.  To compute the number of detectable 
planets, we must specify both the frequency and properties of the planets, 
as well as the the experimental parameters for each technique. 

For most of the results presented in this paper, we consider only
planets of mass $M=\me$ (or radius $R=R_\oplus$).  This is because all
the currently known terrestrial planets\footnote{Around main-sequence
stars.} have $M\le \me$.  However, current theories of planet
formation predict that terrestrial planets may have masses of up to
$\sim 10~\me$, and we therefore consider the sensitivity of the
techniques as a function of planet mass in \S\ref{sec:mass}.  In order
to be as democratic as possible, we begin by assuming orbital periods
that allow each method to play to its strengths.  For astrometry, RV,
and microlensing, we assume that the planets are all located at
periods at which the detection sensitivity of the particular method is
maximized (or nearly so).  For transits, we assume the planets are
located in the habitable zones of their parent stars. Our specific
choices are somewhat arbitrary, and in \S\ref{sec:period} we explore
the sensitivity of the various methods as a function of orbital
period.  Finally, we assume that every main-sequence star surveyed
has a planet of the specified parameters.

We specify the experimental parameters for each technique that
are appropriate to either planned space missions, or future 
ground-based surveys of our
own conception.  Again, our choices here are somewhat arbitrary, and
one could consider other choices for the experimental parameters.
However, we consider our choices to represent the
most ambitious yet feasible experiments of their kind, and therefore
to be very optimistic.

\medskip
\noindent
{\em Radial Velocity}: We assume that each star has a planet
with $M=\me$ and period $P=0.5\,\rm yr$.  We assume a total observing
time and photon collection rate appropriate to four dedicated 2m
telescopes over 5 years.  This corresponds to 60,000 hours of
dedicated observations.  We assume a systematics limit of
$\sigma=1\,\rm m\,s^{-1}$ and take the error to be $\sigma=(1 +
10^{0.4[V-1.5]})^{1/2}\,\rm m\,s^{-1}$ for a 2 minute exposure.  Note
that, in terms total photon collecting power, this is approximately
equivalent to 500 hours per year for 5 years on Keck.  However, the
combination of four dedicated 2m telescopes is much superior to the
Keck scenario for stars $V<5$ because the fixed overhead time per
target is a much smaller percentage of the exposure time.

\medskip
\noindent
{\em Astrometry}: We assume that
each star has a planet with $M=\me$ and $P=3\,\rm yr$.
We assume a survey of 40,000 astrometric measurements each with 
systematics-limited errors of $\sigma=1\,\muas$.  This corresponds to
2,000 hours of {\it SIM} time per year over 5 years, with each
measurement taking 15 minutes.  To account for the photon noise of 
faint stars, we assume an error $\sigma=(1 + 10^{0.4[V-11]})^{1/2}\muas$.

\medskip
\noindent
{\em Transits}: We assume each star has a planet with
$R=R_\oplus$ located in the habitable zone, defined as
$a=(L_\bol/L_{\bol,\odot})^{1/2}\au$, where $L_\bol/L_{\bol,\odot}$
is the bolometric luminosity of the parent star relative to the Sun.
We use the parameters of the {\it Kepler} satellite, except that we 
augment the sample to
include faint nearby stars as advocated by \citet{gpd}, thereby
multiplying the detection rate by a factor 2.5 relative to the
original {\it Kepler} strategy.

\medskip
\noindent
{\em Microlensing}: We assume that each 
main-sequence star has an Earth-mass planet at $a=2.5\,\au$.
We use the results
of a simulation by \citet{ghg} of a 5-year ground-based microlensing search
using four 2m wide-field telescopes located in Chile, South Africa, Australia, and Hawaii.  This simulation adopts the star-count
based normalization of the optical depth calculated by \citet{hg}.
The form of the derived S/N distribution is similar to that of the
MPF satellite \citep{gest}, but the normalization is about 30\% smaller
when the latter is rescaled using the same optical depth derived by
\citet{hg}.

For transits and microlensing, the above descriptions serve to
completely describe the experiment, but for astrometry and RV they do
not.  Transits and microlensing are carried out in survey mode, so
that every star is, by definition, observed in exactly the same way.
Astrometry and RV are pointed experiments in which the observer is
free to divide up the observing time among targets with more or less
complete freedom.  In our baseline comparison, we will force
astrometry and RV into survey mode.  That is, we will demand that they
allocate the same amount of observing time to each of their program
stars.  We determine the number of program stars, $N$, by rank
ordering all stars in the {\it Hipparcos} catalog \citep{hip}
according to their sensitivity under each respective method, and then
requiring that a $1\,M_\oplus$ planet in a favorable orbit (edge-on
for RV, face on for astrometry) around the $N$th star can be detected
at $\sn=5$ by using $1/N$ of the total observing time.  This may not be
the optimal choice and is certainly {\it not} how current RV surveys
are carried out.  Hence, in \S~\ref{sec:flex} we also consider alternative
observing scenarios.

Of course, our assumptions are idealized in that they do not take
account of both recognized and unrecognized effects.  For example,
photospheric variations will prevent $\sigma=1\,\rm m\,s^{-1}$
measurements for some RV targets; some reference stars required to fix
the local reference frame of astrometric targets will be found to be
astrometrically unstable; some transit targets will be too variable
for reliable photometric monitoring; we ignore binary stars
completely.  Since problems of this type are likely to affect all methods,
and their actual magnitude cannot be estimated with high precision in
advance, we ignore them here.  As nearly all unaccounted-for 
effects will serve to degrade the sensitivities, our assumptions tend to be optimistic. 

\begin{figure}
\epsscale{1.1}
\plotone{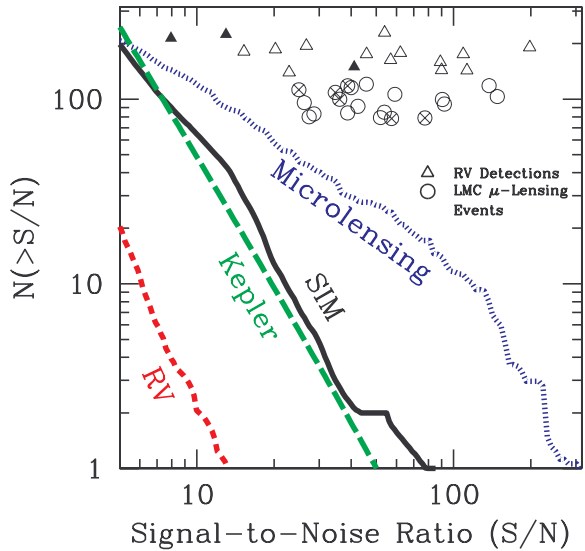}
\caption{\label{fig:one}
Number of detectable Earth-mass planets as a function of threshold in S/N.
Each experiment is allowed to play to its strengths; 
it is assumed that all stars have planets with: $P=0.5\,$yr
for the RV survey;
$P=3\,$yr for astrometry ({\it SIM}); $a=2.5\,\au$ for microlensing;
and $a=(L_\bol/L_{\bol,\odot})^{1/2}\au$ for transits.  Symbols show the actual 
S/N of detections in two challenging historical experiments.  Circles passed
the basic cuts as microlensing events toward the LMC as reported
by \citet{macho57}, but circles with crosses were later found to
be background SNe, not microlensing events.  Triangles are RV detections
of planets.  Filled triangles are detections that are ambiguous,
controversial, or have been retracted.
}\end{figure}

\section{Scaling with Signal-to-Noise Ratio}\label{sec:scaling}

Figure~\ref{fig:one} and Table 1 compare the distribution of 
signal-to-noise ratios of all four methods.  
Here we define the signal-to-noise ratio
as $\sn\equiv (\Delta\chi^2)^{1/2}$, where $\Delta\chi^2$ is the
difference in $\chi^2$ in the best-fitting models with and without a
planet.  The scalings with $\sn$ are all basically power laws
$N\propto (\sn)_\min^{-\nu}$, except for the case of microlensing,
for which the scaling is a broken power law.  Why do the functions take
this form and why is the slope for microlensing so much shallower than
the other methods?  While each curve is the result of a complex
numerical integration, it is nevertheless possible to basically
understand the origins of all four slopes using fairly simple
reasoning.  In each case, we need to ask, given an ensemble of
detections with S/N above some threshold, what subset of these
detections will have twice that S/N?

Let us begin with {\it Kepler} transits.  Consider a star of given
spectral type for which a planet can just barely be detected at
$\sn=(\sn)_\min$.  Where can one find a star of the same spectral
type with $\sn=2(\sn)_\min$?  Clearly, if another star were closer
by a factor 2, it would yield 4 times greater flux and so twice as
much S/N.  However, there are 8 times fewer of such nearby
stars, so $\nu=3$.  Why then is the slope in Figure~\ref{fig:one}
$\nu=2.4$ and not exactly 3?  As discussed by \citet{gpd}, the
{\it Kepler} $10''$ point spread function (PSF) contains about
$V=17$ of sky, which fundamentally limits its sensitivity to otherwise
detectable planets around M dwarfs.  For each adopted threshold
$(\sn)_\min$, there is some spectral class of stars for which
coming closer by a factor 2 moves the star from below the sky
to above it, and this implies that it is not quite necessary
to reduce the sample size by a factor 8 to double $(\sn)_\min$.
If all stars were above the sky, the slope would be $\nu=3$.  If all were below
the sky, the slope would be $\nu=1.5$.  The actual value of $\nu=2.4$ results 
from competition between these two regimes.

The reasoning is quite similar for {\it SIM} astrometry.  In this
case, the astrometric precision does not depend on distance
for most stars that will be monitored because the systematics
limit sets in relatively faint at $V=11$.  Since the astrometric
signature scales inversely as the distance, while the astrometric
noise is distance-independent, the S/N also scales inversely
as the distance.  So, just as with {\it Kepler}, one expects
$\nu=3$.  However, just as with {\it Kepler}, this scaling is
broken by a second-order effect.  Some stars are $V\ga 11$ and
hence are in the photometric not systematics limit.  In the photometric 
limit, one expects $\nu=1.5$.  The net value of $\nu=2.4$ again
results from competition.

For RV, the situation is qualitatively similar.  In the photon limit,
one expects $\nu=3$, just as for {\it Kepler}.  Although very few stars
have $V<1.5$ (the regime dominated  by systematics), some are this bright.  Once
systematics set in, there is no improvement in S/N obtained by moving
the star closer.  Hence, $\nu=\infty$.  The net result is that
$\nu=3.1$, i.e., slightly higher than the value obtained in the photon
limit.

For microlensing, the situation is very different from the other three
methods.  First, the sources of light are independent of the planets
and their hosts, and furthermore, are all at essentially the same distance
(i.e., the Galactocentric distance $R_0\sim 8~{\rm kpc}$).  Therefore,
the distribution of photometric errors directly reflects the shape of
the source-star luminosity function.  Second, there is large
distribution of signal amplitudes $\delta$ for a given planetary
system.  Therefore, in order to find a higher $\sn$ event, one can
either look at a brighter source stars, or ``wait'' for the rare events
that have higher intrinsic signal amplitudes.  
It is possible to demonstrate numerically that,
for a planet with mass ratio $q$ located near the Einstein ring of its
primary, the area of the Einstein ring that is photometrically
perturbed by the planet by a fractional amount $>\delta$ is $\Omega
\propto q\delta^{-1}$.  This is in the limit of point sources; in reality
this distribution will be cut off at a some maximum deviation $\delta_{\rm max}$
due to finite-source effects. The luminosity function in the Galactic bulge 
is well-represented by a broken power-law, with
${\rm d}N/{\rm d}{\rm log}{L} \propto L^{-1/3}$ for stars fainter than the turn-off, and
$\propto L^{-1.3}$ for stars brighter than the turn-off
\citep{holtzman98}.  Stars fainter than the turn-off are fainter than
the combined sky plus unresolved star background, and thus $\sn \propto
L \Omega^{1/2} \delta \propto L q^{1/4} \delta^{3/4}$, where the
$\Omega^{1/2}$ term arises from the assumption that the typical
duration of perturbations is $\propto \Omega^{1/2}$.  Stars brighter
than the turnoff are typically above the background, and thus $\sn
\propto L^{1/2} q^{1/4} \delta^{3/4}$.  These basic ingredients 
combine to yield the broken power-law behavior in seen in 
Figure~\ref{fig:one}.  We
note that, for a space-based mission such as {\it MPF}, essentially
all sources are above the background.  Therefore, for sources below
the turn-off the scaling with $\sn$ is expected to be
steeper than for the ground-based scenario (although the overall normalization is higher), as is seen in the results of the simulations of \citet{gest}.

\bigskip
\bigskip
\begin{table*}[t]
\begin{center}
{\large TABLE 1}\\
\smallskip
{\sc Number of Detectable Planets}\\
\smallskip
\begin{tabular}{c|ccccc}
\hline
Method & $\sn>5$ & $\sn>10$ & $\sn>25$ & $\sn>50$ &$\sn>100$ \\
\hline
\hline
RV (Survey)    &  22  &   2  &  0  &  0  &  0 \\
SIM (Survey)   & 207  &  65  &  7  &  2  &  0 \\
Microlensing   & 208  &  121  & 46  & 26  & 11 \\
Kepler         & 246  &  49  &  6  &  1  &  0 \\
\hline
 	& $\sn=5$ & $\sn=10$ & $\sn=25$ & $\sn=50$ &$\sn=100$ \\
\hline
\hline
RV  (Optimized)  &  38  &  15  &  5  &  2  &  0 \\
SIM  (Optimized) & 335  & 174  & 76  & 37  & 17 \\
\hline
\hline
\end{tabular}
\end{center}
\end{table*}

\subsection{Flexibility of Astrometry and RV}\label{sec:flex}

\begin{figure}
\epsscale{1.1}
\plotone{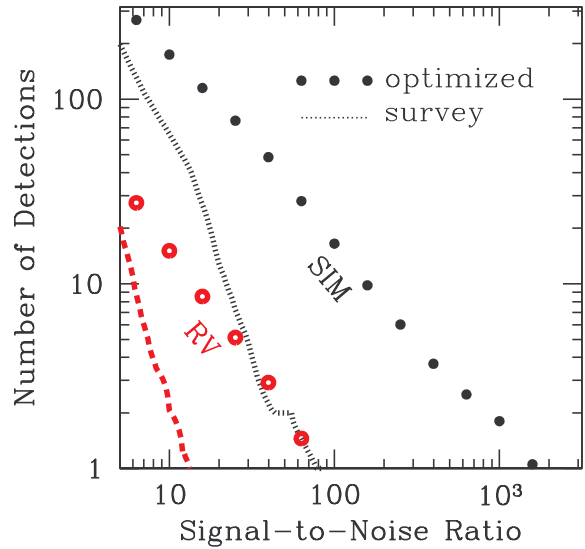}
\caption{\label{fig:two}
Flexibility of RV and astrometry ({\it SIM}).  Curves are the
same as Fig.~\ref{fig:one} and represent the sensitivity of these
techniques if they are forced into ``survey mode'', in which each target
star is observed for the same total amount of time.  Dashed curve
is for RV, dotted curve is for {\it SIM}.  Circles represent
a series of {\it different experiments}, in each of which the exposure
times are tailored to obtain the same S/N for each target star.  That is,
there are more detections at a fixed S/N but at the price of not having
a tail of detections at substantially higher S/N.  Filled
circles are for {\it SIM}, open circles are for RV.
}\end{figure}

Recall that in constructing Figure~\ref{fig:one}, we handicapped
the astrometric and RV experiments to mimic surveys.  
That is, we required them to expend the same observing time on all targets.
The transit and microlensing surveys must do this, but in reality
astrometry and RV are not so constrained.  
We illustrate this flexibility by considering an alternative set
of strategies.  Here we first choose an arbitrary number $N$ and 
pick out the $N$ stars with the best sensitivities from our rank-ordered
list of {\it Hipparcos} stars.  We divide all the available observing
time among these $N$ stars in such a way that each has equal final
S/N.  That is, the worst stars from among the list are observed more
frequently (or with longer exposures) than the best.   The result
of this exercise is shown by the chains of filled
circles in Figure~\ref{fig:two},
which should be contrasted with the corresponding curves taken
from Figure~\ref{fig:one}.   Table 1 compares the number of detected
planets at or above several selected $\sn$   
under these two different observational
strategies.  
It is important to emphasize that the curves represent one experiment,
while the circles represent many different possible experiments,
only one of which can be realized.  That is, if every star has a planet,
and the astrometric observations follow the survey strategy, then there
will be three detections with $\sn>40$ and 30 detections with $\sn>16$.
However, if the alternative strategy is used, one has a choice of
three detections with $\sn=500$ or 30 detections with $\sn=60$.  One
cannot have both.

\section{Conservative Signal-to-Noise Thresholds}\label{sec:thresholds}

What is the appropriate S/N cutoff?  The answer to this question
depends on one's definition of ``appropriate.''  Traditionally, the
appropriate value of the minimum S/N threshold has been set by
just considering statistics.  One can ask, given the expected
properties of the noise and signal, what S/N threshold is necessary in
order that less than some tolerated number of false alarms are
expected.  Determination of this ``statistical'' $(\sn)_\min$ can be
quite complicated and requires careful consideration of both the
expected distribution of measurement noise as well as the number of
statistically independent trials needed to thoroughly search the data (see, e.g.,
\citealt{jenkins02}).

Thresholds set based on statistics alone are the least conservative
and the most optimistic, in the sense that it is statistically
impossible to detect a signal with any confidence at a lower
threshold.  While such thresholds maximize the number of detected
planets, we argue that they are dangerous and undesirable for several
reasons.  First, for the majority of the ``detections'', which are near
this minimum threshold, one has used essentially all of the
available information simply to detect the signal.  This means that it
is generally impossible to determine the nature of the signal, i.e.,
to measure planetary parameters or to use the detailed form of the
signal to corroborate its interpretation.  Simply speaking, all one is
confident of is that {\it something} has been detected -- there is
essentially no information about exactly {\it what} has been detected.
As a result, such detections are extremely prone to being confused
with false positives.  Given that, for many of these experiments,
follow-up and independent confirmation of the candidates will be difficult or impossible, this is a serious
concern.  We discuss this in more detail below. 
Furthermore, even if false positives can be reliably
excluded, it is still the case that the parameters of the detected
planets will be very poorly constrained, making the scientific
usefulness of the detections questionable.

We argue that the actual S/N threshold adopted should
consider both the ability to extract physical parameters from the
detected planets, as well as the possible presence of false positives,
and the inability to perform follow-up of the detections.  The
threshold required to measure planetary parameters to a given
precision depends on many different quantities, such as the complexity
of the signal, the cadence, sampling, and duration of the
observations, etc.  This issue has been most thoroughly explored in
the case of {\it SIM} \citep{sozzetti02,sozzetti03,ft03}.  Very
roughly, these authors find that thresholds of $\sn \ga 25$ are
required to constrain the planetary parameters to better than $30\%$.
For transits, the situation is considerably better, as the period of
transiting planets should be measured with exquisite precision even
for low $\sn$ detections, and the fractional error in the planetary
radius is $\sim 0.5(\sn)^{-1} \la 2\%$ for $\sn > 25$.  For
microlensing, the simulations of \citet{ghg} indicate that, for $90\%$
of the perturbations detected with $\sn>25$, the planet/star mass
ratio should be measurable to $\la 30\%$.  No comprehensive study of the 
ability to extract planet parameters from RV curves has been performed, but
it seems likely that, given the nature of the observations, the uncertainties will be similar to those for astrometric data.  

What $\sn$ threshold is required to deal with false positives?
Until the experiments are actually carried out, and false positives
are identified, one cannot be certain.  
However, it is instructive to consider historical precedent.
Figure~\ref{fig:one} summarizes two relevant experiences.
The first is the 5.7-year microlensing dark-matter search toward the
Large Magellanic Cloud (LMC) by the MACHO collaboration \citep{macho57}.
Like the proposed planet searches, the MACHO experiment was a massive
search for objects that are extremely difficult to detect.  Because the
detections involve events that are over, and hence can {\it never}
be verified individually, MACHO was compelled to be very conservative
in what they called a ``detection''.  They demanded $(\sn)_\min=20$.
Nevertheless, there is substantial debate in the microlensing
community as to whether they were conservative enough.  Figure~\ref{fig:one}
illustrates one reason why.  The circles show the S/N of all the events
surviving the basic selection cuts.  However, circles with crosses
were eliminated because they were judged to be probable supernovae (SNe).
How they came to be recognized as such is very instructive in the current
context.  Originally, SNe were not considered as a possible background.
It was only because two of the SNe ``events'' had very high $\sn\sim 60$
(see Fig.~\ref{fig:one}) that their character as SNe was recognizable.
Then other, lower S/N, events were examined for telltale SN signatures.
If all the events had had $\sn<40$, it is unlikely the SNe would have
ever been recognized.  For this reason, one may also wonder whether
some of the other ``low-quality'' ($\sn\sim 30$!) events are due to
some other unanticipated astrophysical effect.  \citet{macho57} were
not able to prove that they were not.  However, because of the strong
tail of very high S/N events, they were able to show that their 
conclusions did not rest on these events that were open to question:
they repeated their analysis, progressively increasing $(\sn)_\min$,
and showed that their conclusions did not change (although of course
their statistical errors then deteriorated).  The high S/N tail
therefore gave credence to lower S/N events, at least statistically,
and these then allowed MACHO to better characterize the population
they were detecting.

Also shown in Figure~\ref{fig:one} is a (non-exhaustive) sample of RV
planet detections from \citet{cumming99} and \citet{vogt02}.  Note
that the great majority of these have very high S/N.  This partly
reflects the extreme conservatism of the observers.  This conservatism
is practiced by the majority of the RV observational community and is
reflected in the fact that, while the typically quoted
single-measurement RV error is only a few ${\rm m~s^{-1}}$, very few
planets have been announced with velocity semi-amplitude $\la 20~{\rm
m~s^{-1}}$.  It is just such conservatism that has established the RV
planets as bedrock facts rather than hopeful speculation.  The
RV detections of planets around HD219542b and $\epsilon$ Eri
are the exceptions that prove the rule.  The tentative detection of a
planet around HD219542b was originally claimed by \citet{desidera03}.  This
detection was at $\sn \sim 8$, near the $\sn$ threshold proposed for
terrestrial planet searches.  However, \citet{desidera03} cautioned
that the RV variations could also be due to stellar activity.  Indeed,
\citet{desidera04} recently reported additional measurements
indicating that the RV variations are indeed most likely due to stellar
activity.  The RV
detection of a planet around $\epsilon$ Eri was at an even higher
$\sn\sim 13$ \citep{epseri}.  However, despite this
relatively high S/N (by the standards of proposed terrestrial planet
searches) and despite the strong external prior favoring a planet
around $\epsilon$ Eri \citep{notspock,spock}, this detection remains
controversial.  Also very instructive is the extremely strong
($\sn\sim 40$) detection reported for HD192263 \citep{santos00}, which
was later refuted \citep{henry02} and then subsequently reasserted
\citep{santos03}.  All three of these examples show that extreme caution is
warranted when probing for extreme or novel phenomena.

The microlensing-like SNe, HD219542b, and $\epsilon$ Eri demonstrate that it is
difficult to break new scientific ground with detections that, while
formally robust, lack the S/N to unambiguously characterize the object
that has been detected.  On the contrary, the significance of such
ambiguous ``detections'' can only be evaluated provided that the
sample includes other detections that are completely unambiguous.

Now, it is true that experiments yielding null results can and do
operate at or near the pure-noise limit.  For example, \citet{47tuc}
was able to set limits on planets in 47 Tuc going down to a threshold
$(\sn)_\min=6.3$ at which point they expected $O(1)$ false detection
due to Gaussian noise. \citet{muplanets} searched for planetary
signatures in 43 microlensing events down to $(\sn)_\min=7.7$, a
factor 1.5 higher than the threshold dictated by Gaussian noise.
Their retreat from this threshold is highly instructive.  They found
about a half dozen events above the Gaussian threshold but below their
adopted one.  They concluded that these could not be regarded as
reliable planet detections partly because a Monte Carlo analysis of
stable light curves showed a similar S/N distribution, and partly
because of the expected form of the microlensing S/N distribution
shown in Figure~\ref{fig:one}.  That is, if a substantial fraction of
these marginal ``detections'' were real, one would also expect other
detections at much higher S/N.  This again demonstrates that without
good sensitivity to high S/N detections, a cutting-edge experiment
cannot reliably interpret the marginal detections.

\section{Results}\label{sec:results}

\subsection{Number of Planets at Realistic Thresholds}\label{sec:numvthresh}

Considering all the arguments presented in \S\ref{sec:thresholds}, 
we conclude that, although
thresholds of $(\sn)_\min \sim 8$ may yield formally statistically
significant detections, conservative
thresholds of $(\sn)_\min \sim 25$ should be adopted for the ambitious,
cutting-edge experiments discussed here. In light of this conclusion,
it is instructive to reconsider the signal-to-noise scalings show in
Figure \ref{fig:one} and discussed in \S\ref{sec:scaling}.  Table 1
summarizes the number of expected Earth-mass planet
detections for various values of $(\sn)_\min$.  Several important
points are worth mentioning. First, although transits ({\it Kepler}),
astrometry ({\it SIM}), and microlensing all have similar sensitivities
when compared at a minimal requirement $(\sn)_\min = 8$, the
sensitivities of both {\it SIM} and {\it Kepler} fall off very steeply
compared to microlensing ($N\propto (\sn)_\min^{-2.4}$ versus
$N\propto (\sn)_\min^{-1}$).  As a result, at the conservative threshold
of $(\sn)_\min=25$, microlensing is 5 times more sensitive than either
transits or astrometry.  Furthermore, for the specific assumptions
(see \S\ref{sec:methods}) adopted here, the total number of Earth-mass
planets detected by {\it Kepler} and {\it SIM} above this threshold is
relatively small, $\la 10$.  In the case of {\it SIM}, this may be
improved by more judicious appropriation of the observing time
(\S\ref{sec:flex}), and the situation will be better for shorter
periods (for {\it Kepler}) or longer periods (for {\it SIM}).
However, it is generally the case that the assumptions we have made
are optimistic, as we have ignored known (and unanticipated)
effects that will degrade the sensitivity, assumed that {\it all}
stars have Earth-mass planets, and assumed
observational strategies that are more aggressive than those currently being planned for
these experiments.  Therefore, it seems likely that, when more
realistic assumptions are adopted, the number of expected robust
detection of Earth-mass planets will be modest.

\begin{figure}
\epsscale{1.1}
\plotone{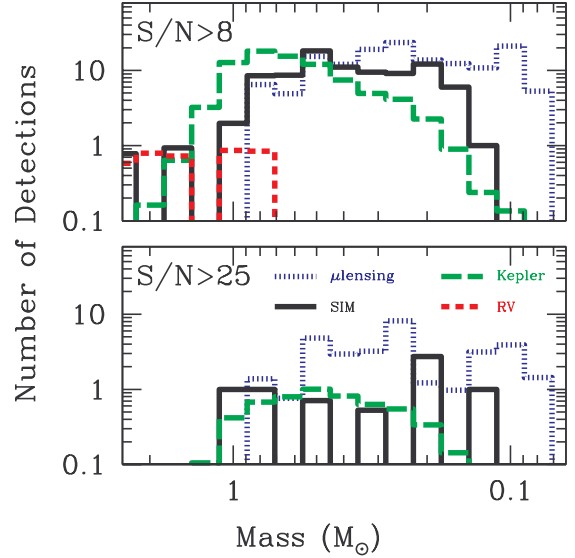}
\caption{\label{fig:three}
Distributions of parent-star masses for the experiments summarized
in Fig.~\ref{fig:one}, for two different S/N thresholds, 
$(\sn)_\min=8$ and $(\sn)_\min=25$.  At $(\sn)_\min=8$, microlensing,
transits ({\it Kepler}), and astrometry ({\it SIM}) all have
comparably broad parent-star mass distributions with comparable
normalizations.  At $(\sn)_\min=25$, the forms of the distributions
remain similar, but microlensing has a much higher normalization
than the other two.
}\end{figure}

\subsection{Sensitivity as a Function of Stellar Mass}\label{sec:comp}

What are the masses of the stars whose planets would be detected
in the four surveys summarized in Figure~\ref{fig:one}?
Figure~\ref{fig:three} shows these distributions for two
different S/N thresholds, $(\sn)_\min=8$ and $(\sn)_\min=25$.

At  $(\sn)_\min=8$, which is similar to the thresholds that are
customarily discussed in this field,  microlensing, astrometry,
and transits all have rather similar distributions, with microlensing
extending to somewhat lower masses and transits to somewhat higher
masses.  These broad distributions primarily reflect the breadth
of the underlying stellar mass function.  The {\it Kepler} experiment
tends to be cut off at low masses because these faint stars fall
below the sky of its $10''$ PSF \citep{gpd}.  Microlensing is cut off
at high masses because there are no such stars in the Galactic bulge,
which is the location of most of the lens population.  RV is especially
sensitive to high-mass stars because these are bright and, except
for stars $V\la 1.5$, our assumed sensitivity scales  
$\propto 10^{-0.2 V}$.  In fact, the most massive stars in the RV
histogram are A stars and therefore probably inaccessible to RV.
This subtlety has been ignored in our simple prescription for detectability
based on $V$ flux alone.

However, as we discussed in \S~\ref{sec:thresholds}, historical experience
argues against the viability and believability of $\sn=8$ detections
in challenging experiments of this type.  Based on this historical 
experience, we believe that the experimental design should hinge on the
expected number and distribution of detections at much higher S/N, 
perhaps $(\sn)_\min=25$.  This is also shown in Figure~\ref{fig:three}.
The functional forms of the distributions are similar to the
case of $(\sn)_\min=8$, but the amplitudes are reduced.  As
expected from Figure~\ref{fig:one} and the scaling relations
discussed in \S~\ref{sec:scaling}, microlensing drops by about a factor of 
3, while astrometry and transits drop by about a factor 15.  RV would
not detect Earth-mass planets at all at this threshold.

\subsection{Sensitivity as a Function of Planet Period}\label{sec:period}

\begin{figure}
\epsscale{1.1}
\plotone{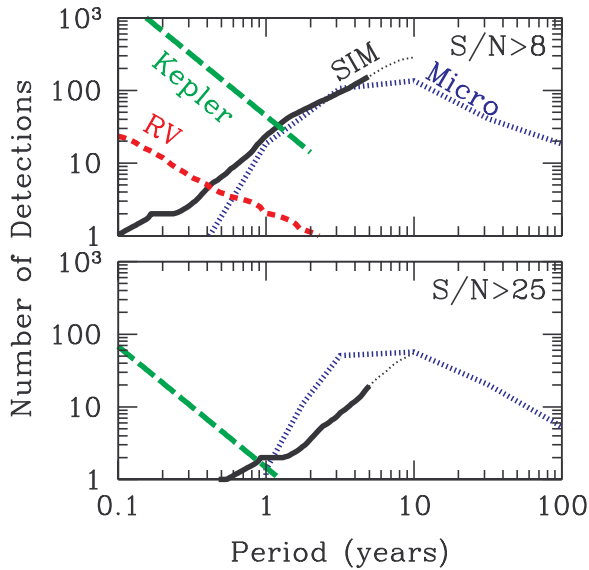}
\caption{\label{fig:four}
Sensitivity to planets as a function of period $P$ for the four
techniques.  In contrast to Figs~\ref{fig:one}--\ref{fig:three}, here
it is assumed that every star has a planet with period $P$, so
that each technique is confronted with the same ensemble of planetary
systems.  Transits ({\it Kepler}) are truncated at $P=2\,$yr because
{\it Kepler} is a four-year experiment that must detect at least 2
transits.  Astrometry ({\it SIM}) has a dotted-line extension from
$P=5\,$yr to $P=10\,$yr to reflect the possible extension of this
mission from 5 to 10 years.  The survey designs are identical to
those described in \S~\ref{sec:complement}, and the indicated
S/N thresholds are the same as in Fig.~\ref{fig:three}.
}\end{figure}

As discussed in \S~\ref{sec:complement}, our primary comparison
of different techniques has allowed each to play to its strengths.
While democratic, this approach also means that 
Figures~\ref{fig:one}--\ref{fig:three} do not reflect the relative
detectability of the same ensemble of planetary systems.  To
better understand the complementarity of different techniques
that was discussed in \S~\ref{sec:complement}, we plot in
Figure~\ref{fig:four} the number of detectable planets as a function
of period, $P$.  In this case, we assume that each star has one
planet at each period.  As expected, transits and RV are primarily
sensitive to short-period planets, while microlensing and astrometry
are primarily sensitive to long-period planets.

\subsection{Sensitivity as a Function of Planet Mass}\label{sec:mass}

For the majority of the discussion, we have concentrated on planets
with $M=\me$ (or $R=R_\oplus$), since all the currently-known
terrestrial planets have a mass $\la \me$. However, some theories of planet
formation predict that considerably more massive terrestrial planets
with $M\la 10~\me$ may be common (e.g., \citealt{ki02,raymond04,il04}). Therefore we
briefly consider the number of expected planet detections as a function of the
mass of the planet.  For both RV and astrometry, the signal is
directly proportional to the mass of the planet.   When combined with
the $\sn$ scalings from \S\ref{sec:scaling}, this yields $N\propto M^{2.4}$ for
astrometry and $N\propto M^{3.1}$ for RV.  For {\it Kepler}, the signal
is $\propto R^2$, or, assuming constant planet densities, $\propto
M^{2/3}$.  Therefore, $N \propto M^{1.6}$.  Finally, for
microlensing, the signal is $\propto M^{1/4}$.  However, here we must
also consider the increase in the detection probability for the larger
mass ratios, which is $\propto M^{1/2}$.  This yields, for sources
below the turnoff and fainter than the background, $N\propto M^{3/4}$.  We confirm this scaling in our simulations.

\section{Discussion}\label{sec:discuss}

As discussed in \S~\ref{sec:complement}, and demonstrated in
Figures~\ref{fig:one} and \ref{fig:four}, the four detection
techniques reviewed here are very distinct, not only in terms of their
sensitivity as a function of $\sn$ and period, but also in the
information they yield about the planets.  In addition, there are
number of secondary characteristics of the planet sample gathered
by the experiments that are
unique to each of the four methods.  These characteristics may also be
important to consider when assessing how well-suited a given method is
to fulfilling a particular task.   In the following, we
examine the relative strengths and weakness of each method, considering
both the primary and secondary characteristics of the planet harvest, and
concentrating on five interrelated considerations: the total number of
planets detected, the sensitivity to habitable planets, the ability to
robustly calibrate the frequency of terrestrial planets as required
for the success of {\it TPF}, the location and nature of the
planet-bearing stars, and the opportunity for follow-up of the
detected planets and hosts.

The RV planet search envisioned here is clearly the least competitive
in terms of the total number of planets detected.  Even an optimized
search using 4 dedicated 2m telescopes searching for 5 years would
only allow the detection (at $\sn=25$) of Earth-mass planets with
$P=0.5~{\rm yr}$ around the five best candidate host stars.  The
sensitivity is further reduced for planets in the habitable zone.  On
the other hand, the technology required to initiate such a search is
already available, and the detection of even one such planet would be
very interesting, because the host stars would be very nearby $d\la
20~{\rm pc}$, and thus promising targets for detailed follow-up with
ground-based instruments, as well as {\it TPF}.  Furthermore, the fact
that such a search is ground-based and targeted means that it is very
flexible.  Therefore, it may be possible to search for Earth-mass
planets around a larger number of stars using a two-tier survey
approach, in which promising targets from an initial, low-$\sn$ survey
are followed up more intensively to yield high-$\sn$ detections.
Finally, RV searches are advantageous in that they can detect
multi-planet systems, with both gas-giant and terrestrial planets,
providing essential constraints on the effects of massive planets on
terrestrial planet formation.  Furthermore, planets detected via RV
can be confirmed by {\it SIM}, yielding planet masses.

In `survey' mode, in which all targets are allocated the same amount
of observing time, and for conservative thresholds of $(\sn)= 25$,
{\it SIM} will detect at most $\sim 10$ Earth-mass planets at
$P=3~{\rm yr}$.  For the largest detectable period corresponding to
the nominal mission lifetime of $P=5~{\rm yr}$, $\sim 30$ Earth-mass
planets are detectable.  However, it may be possible to detect as many
as $\sim 80$ Earth-mass planets at $\sn=25$ and $P=3~{\rm yr}$ if the
search is optimized, and the observing time is allocated according to
the sensitivity of the host star to planets (see \S\ref{sec:flex}).
As discussed above in the context of RV searches, it may be possible
to survey an even larger number of stars for Earth-mass planets by
adopting a dynamic, two-tier approach.  {\it SIM} will also be
sensitive to planets in the habitable zone of some target stars
\citep{sozzetti02}, although maximizing the number of detections of
such planets requires a modification of the observing strategy
\citep{gff03}.  In
addition, the stars surveyed by {\it SIM} are exactly those which can
be targeted by {\it TPF}, making the results of the {\it SIM} planet
search critical.  However, we emphasize that the results presented
here assume a very ambitious {\it SIM} planet search.  In particular,
we have assumed that a total of $\sim 10,000$ hours, or $\sim 20\%$ of the
available {\it SIM} time, is devoted to searching for planets.
Currently only $\sim 8\%$ of {\it SIM} time is allocated to searching
for planets around nearby main-sequence stars, in two {\it SIM} key projects.  Furthermore, the two key projects are not currently
optimized to find Earth-mass planets.  Therefore, achieving the
results presented here will require not only an increase in the amount
of time alloted to planet searches, but also a careful reassessment of
the observing strategies.

The primary advantage of {\it Kepler} is that it will be sensitive to
Earth-mass planets in the habitable zone.  Unfortunately, at
conservative thresholds of $\sn\sim 25$, {\it Kepler} will only find
$\sim 6$ Earth-mass habitable planets.  Although the detection of even
one such planet would be extremely interesting, this is an
insufficient number to accurately constrain the frequency of such
planets.  Therefore, constraints on this frequency will have to be
derived from lower $\sn$ observations, or from extrapolations from the
frequency of higher-mass planets.  Both of these procedures are
potentially risky.  {\it Kepler} is much more sensitive to planets
with shorter periods (see Figure \ref{fig:four}), and should provide
robust statistics on terrestrial-mass planets close to their parent
stars.  Follow-up of candidate Earth-mass planets
detected by {\it Kepler} will be difficult.  
The target stars of {\it Kepler} are
disk main-sequence stars at distances of $\sim \kpc$, and thus will be
too faint ($V>9$) or distant for confirmation with either RV or {\it
SIM}.  Naively, one might expect the Hubble Space Telescope ({\it HST}),
provided that it is still operational, to yield $\sn$ that are larger
by a factor of two, given that its aperture is twice that of {\it
Kepler}.  However, this is more than compensated for by the fact that
{\it Kepler}'s bandpass is larger by a factor of $\sim 10$ than
the $50~{\rm nm}$ bandpass used to
observe the transiting planet HD209458b, which \citet{brown01}
found to yield photon rates that are a
factor of $\sim 5$ smaller than those expected for {\it Kepler} for
stars of the same magnitude.  Therefore, one expects $\sn$ that are a
factor of $\sim 2$ smaller for {\it HST}.  It may be possible to
follow-up candidate transit events in the infrared using the James Webb Space
Telescope ({\it JWST}). This will depend on the photometric precision
of the available {\it JWST} instruments and on whether it is launched
in time to confirm the {\it Kepler} transits before their phases are
lost.

Of the four methods we have discussed, microlensing is the most
sensitive in terms of the total number of potential planet detections
at realistic $\sn$ thresholds.  At $\sn \sim 25$, a 5-year
ground-based microlensing survey of the kind envisioned here could
potentially detect $\sim 50$ Earth-mass planets at $a=2.5~\au$.
Therefore, microlensing could constrain the frequency of terrestrial
planets to reasonable precision.  Although such a survey would also be
sensitive to Earth-mass planets at $\sim 1\au$, this is only for
low-mass target stars, and thus microlensing is not able to directly
constrain the frequency of Earth-mass planets in the habitable zone.
While a ground-based microlensing survey is attractive in that it is
flexible, relatively inexpensive, and does not rely on untested
technology, the results stated above are predicated on the ability to
achieve near photon-limited photometry on stars with $I\ga 20$
\citep{ghg}.  This may be difficult in the crowded fields toward the
Galactic bulge \citep{bennett04}.  A space-based microlensing survey
such as {\it MPF} would circumvent the problems with crowded-field
photometry. Detailed simulations of a space-based microlensing survey
indicate that it would yield a detection rate of Earth-mass planets at
$\sim 2.5\au$ that is higher than the ground-based survey by at least
$\sim 30\%$ \citep{gest}.  Furthermore, the space-based survey would
be sensitive to planets over a wider range of semi-major axes, and
would be considerably more sensitive to multiple planets.  Both
ground- and spaced-based microlensing surveys are sensitive to
Earth-mass planets surrounding main-sequence stars with distances of
$1-10~\kpc$.  It is important to emphasize that microlensing is
sensitive to planets around both Galactic bulge and disk stars.  For
example, in the simulations of \citet{ghg}, $\sim 30\%$ of the
detected Earth-mass planets orbit disk main-sequence stars.  Because
of the large distances to the lens stars, follow-up of detected planet
candidates will be difficult.  However, in many cases the lens star
should be sufficiently bright to be detectable, and it will be
possible to constrain the mass of the primary, and thus the mass of
the planet \citep{gest}.  Furthermore, by combining ground and
space-based surveys, it should be possible to routinely measure planet
masses \citep{mulensmass}.  However, direct confirmation of the planet
candidates via, e.g.\ RV or astrometric follow-up, will be essentially
impossible in the foreseeable future.

The synergy of the four methods for detecting Earth-mass planets is
clear from Figure~\ref{fig:four}, which demonstrates the possibility
of searching for Earth-mass planets over many decades in period by
combining a variety of techniques.  Since nothing whatsoever is known
about the distribution or frequency of Earth-mass planets around other
stars, no technique can be guaranteed to be the most sensitive.  Any
technique could yield null results if the regions of its greatest
sensitivity were sparsely populated by planets.  Therefore, a combined
attack on the problem using all four techniques is indicated.  The
discussions above indicate that the result of such a coordinated
search for planets could go a long way toward answering many of the
questions of interest, and would provide important calibrations for the
next generation of experiments, including direct searches like {\it
TPF} and {\it Darwin}.

That said, the results presented here, and in particular as shown in
Figure~\ref{fig:four}, also reveal some rather sobering realities.
The combined sensitivity of the four techniques reaches its nadir at
$P\sim 1\,$yr, near the periods of the only two Earth-mass planets
known to be orbiting a main-sequence star.  Furthermore, at the
$(\sn)_\min=25$ threshold that historical experience suggests is the
minimum for reliable results on such cutting-edge experiments, the
number of expected detections of Earth-mass planets with $P \sim
1\,{\rm yr}$ is frighteningly low.  Even if all stars had such planets
and even if all four of these techniques were funded and/or allocated
telescope time at the level we have somewhat optimistically supposed,
and even if all worked according to their specifications, only about 5
such planets would be detected at $(\sn)_\min=25$.  Indeed, unless
Earth-mass planets are very common, or unless they are packed much
closer to their parent stars than is true in the solar system, then it
is possible that neither {\it Kepler} nor {\it SIM} (as they are
currently planned) would yield any reliable, high-$\sn$ Earth-mass
planet detections.

\acknowledgments 
Work by AG was supported by JPL contract 1226901 and NSF grant 02-01266.
Work by BSG was supported by a Menzel Fellowship from
the Harvard College Observatory.  Work 
by CH was supported by Astrophysical Research Center for
the Structure and Evolution of the Cosmos (ARCSEC) of the Korean
Science \& Engineering Foundation (KOSEF), through the Science
Research (SRC) program, and by JPL contract 1226901.

\clearpage

\end{document}